\documentclass[aps,pra,twocolumn,groupedaddress]{revtex4-1}

\usepackage{graphicx}	
\usepackage{amssymb}
\usepackage{amsmath}
\usepackage{bbm}

\begin{document}

\title{Modeling coherent errors in quantum error correction}

\author{Daniel Greenbaum}
\email[]{greenbaum.daniel@gmail.com}
\author{Zachary Dutton}
%\email[]{zac.dutton@raytheon.com}
\affiliation{Quantum Information Processing Group, Raytheon BBN Technologies, Cambridge, Massachusetts 02138, USA}

\date{\today}

\begin{abstract}
Analysis of quantum error correcting codes is typically done using a stochastic, Pauli channel error model for describing the noise on physical qubits. However, it was recently found that coherent errors (systematic rotations) on physical data qubits result in both physical and logical error rates that differ significantly from those predicted by a Pauli model. Here we examine the accuracy of the Pauli approximation for coherent errors on data qubits under the repetition code. We analytically evaluate the logical error as a function of concatenation level and code distance. We find that coherent errors result in logical errors that are partially coherent and therefore non-Pauli. However, the coherent part of the error is negligible after two or more concatenation levels or at fewer than $\epsilon^{-(d-1)}$ error correction cycles, where $\epsilon \ll 1$ is the rotation angle error per cycle for a single physical qubit and $d$ is the code distance. These results lend support to the validity of modeling coherent errors using a Pauli channel under some minimum requirements for code distance and/or concatenation.
\end{abstract}

%\keywords{}

\maketitle


\section{Introduction}
Progress in fault-tolerant quantum computation relies on the ability to simulate the performance of quantum error correcting codes. For example, the numerical prediction of a high fault-tolerant error threshold for the surface code~\cite{wang_surface_2011} is one of the motivating factors in the significant recent experimental effort to realize topological codes~\cite{corcoles_demonstration_2015,barends_superconducting_2014,nigg_quantum_2014}. Numerical predictions of performance metrics such as the fault tolerant threshold and the logical failure rate typically assume a stochastic (incoherent) and uncorrelated Pauli channel model for physical qubit errors, since this model is easiest to simulate. However, recent findings indicate that a Pauli channel significantly underestimates the diamond norm error rate of coherent errors - errors that are both unitary and slowly varying relative to the gate time~\cite{ball_effect_2016,kueng_comparing_2016,wallman_bounding_2015,wallman_estimating_2015}. Such errors can occur, for example, due to systematic control noise, cross-talk, global external fields, and unwanted qubit-qubit interactions. It is therefore important to examine the accuracy of the Pauli approximation for coherent errors in the context of quantum error correction (QEC).

A variety of results have recently appeared that evaluate the impact of realistic noise on QEC. The numerical work of Refs.~\cite{tomita_low-distance_2014,geller_efficient_2013,gutierrez_comparison_2015} has lent support to using a Pauli model for certain types of incoherent errors. These authors performed simulations of QEC for amplitude and phase damping and the corresponding Pauli-twirl approximations, finding no significant difference in logical error rates. This is consistent with a recent result of Wallman which states that non-unital deviations from Pauli channels (as in amplitude damping) do not significantly impact the error rate~\cite{wallman_bounding_2015}. 

A different result was obtained by Fern, {\it{et al.}}~\cite{fern_generalized_2006} in the case of coherent errors. Using a formalism developed by Rahn, {\it{et al.}}~\cite{rahn_exact_2002} for general noise, these authors found that coherent errors in the physical error channel can lead to coherent errors in the logical channel, as manifested by off-diagonal elements in the superoperators for these channels. For the specific example of the $d=3$ Steane code, Ref.~\cite{fern_generalized_2006} found that an off-diagonal element of order $\epsilon$ in the unencoded error superoperator leads to an encoded (logical) error superoperator with off-diagonals of order $\epsilon^3$ and diagonals of order $\epsilon^4$. This leads to a diamond-distance logical error rate of order $1/\epsilon$ greater than would be obtained by replacing the physical error by its Pauli twirl. The same result was also obtained numerically recently~\cite{gutierrez_errors_2016}.

Another recent paper has reported diamond-distance logical error rates for surface codes up to distance $d=10$ for coherent physical errors~\cite{darmawan_tensor-network_2016}. That work also finds discrepancies between coherent physical errors and their Pauli twirl approximation that are consistent with coherent errors at the logical level.

Despite these insights, it remains a challenge to obtain analytic expressions for the logical error map for general noise as a function of code distance and (for concatenated codes) concatenation. Such information can be useful for determining parameter regimes where a Pauli model is valid. Indeed, Ref.~\cite{fern_generalized_2006} considered general channels, deriving upper bounds on superoperator coefficients for the logical error, but not their actual value except for the $d=3$ Steane code. The results of Ref.~\cite{rahn_exact_2002} were limited to diagonal channels, while Refs.~\cite{gutierrez_errors_2016,darmawan_tensor-network_2016} evaluated the logical noise maps numerically, a technique which does not make explicit the scaling of the logical error parameters with $d$. The latter references also considered coherent and incoherent errors individually but not simultaneously.

Our aim in the present work is to obtain analytic expressions for the logical error map due to a combination of coherent and incoherent physical noise, since both are present in real qubits. We work with the repetition code, which, though not a full quantum code in that it cannot correct both X and Z errors, has the advantage of being analytically tractable and yet nontrivial. Indeed, we find that it reproduces the key features of generic codes, saturating the bounds on error channel parameters under concatenation given in~\cite{fern_generalized_2006}.

Our analysis is restricted to the case of a quantum memory (or of gate-independent errors) and perfect syndrome extraction. Consideration of gate-dependent and syndrome extraction errors is left for future work. We also do not consider coherent leakage errors~\cite{wallman_robust_2016,kueng_comparing_2016} or coherent errors due to residual qubit-qubit interactions. The latter are discussed briefly in Section~\ref{qubit-qubit-section}.

We find that two levels of concatenation are sufficient to eliminate the effect of the coherent error on the logical qubits. We also point out that the coherent contribution to the logical error -- as quantified by the infidelity of the entire quantum computation -- becomes important only after a timescale (number of QEC cycles) $\tau_{\mathrm{coh}}$ that increases exponentially with code distance. Our analysis predicts the same scaling of the failure rate with the error model parameters as one obtains using the diamond norm error metric. However it emphasizes the nature of the error process as it unfolds in time. In particular, the coherent error will not be important at modest code distances for which $\tau_{\mathrm{coh}}$ is longer than the correlation time of the physical error. When this is the case, replacing the physical error by its Pauli-twirl accurately determines the logical error probability for quantum computations of arbitrary length. 

\subsection{Repetition code}
We begin with a brief review of the repetition code. For more details see, e.g.,~\cite{raussendorf_key_2012}. The repetition code on $N$ qubits (code distance $d=N$) is defined by the encoding $|\bar{0}\rangle = |00\ldots0\rangle$, $|\bar{1}\rangle = |11\ldots1\rangle$. The logical $X$ operator, which flips $|\bar{0}\rangle$ to $|\bar{1}\rangle$ and vice versa, is denoted $\bar{X}$ and is equal to
\begin{equation}
\bar{X} = X_1X_2\cdots X_N.\label{Xbar}
\end{equation}
Bit flips ($X$ errors) are detected by measuring the parity of neighboring qubits, which is given by the eigenvalues $(\sigma_1,\sigma_2,\cdots,\sigma_{N-1})$ of the {\em stabilizer} operators $S_1 = Z_1 Z_2$, $S_2 = Z_2 Z_3$, \ldots, $S_{N-1} = Z_{N-1} Z_{N}$. Stabilizer eigenvalues are $\pm 1$ corresponding to even or odd parity, and the set of eigenvalues is called the {\em syndrome}. $N-1$ stabilizers are required to encode a single logical qubit.

When the syndrome is measured, the state is projected onto the subspace of the Hilbert space corresponding to that syndrome. E.g., if the syndrome is $(1,1,\ldots,1)$ then the state after syndrome extraction is in the error-free subspace, known as the {\em codespace}. If on the other hand a faulty syndrome is detected, the error can be corrected by flipping (applying the $X$ operator to) the faulty qubit(s), thereby returning the state to the codespace. We do not pause to discuss the procedure for syndrome measurement since we assume this is done without error. Importantly, we note that the association of a syndrome to a particular error is done in a {\em maximum likelihood} rather than deterministic sense -- multiple errors can have the same syndrome (e.g., $X_1$ and $X_2X_3$ for the $N=3$ code) and we choose the one which is most likely given the syndrome. In this way we minimize the error of the encoded (logical) bits. 

\subsection{Single-qubit errors}
For an $N$-qubit register, we consider the error
\begin{equation}
\mathcal{N}^{(N)} = \mathcal{N}\otimes\mathcal{N}\otimes\cdots\otimes\mathcal{N}, \label{error-full}
\end{equation}
where each $\mathcal{N}$ is a single-qubit error operator. This form of error describes many of the physically relevant noise processes affecting qubits, such as cross-talk, systematic control errors, relaxation, dephasing, and external fields. An important noise source not described by Eq.~(\ref{error-full}) is that due to qubit-qubit interactions. We discuss such processes in Section \ref{qubit-qubit-section}.

We assume the following form for the single-qubit error acting on an arbitrary input state $\rho$ per QEC cycle.
\begin{eqnarray}
\mathcal{N}[\rho] &=&  (1-q)e^{-i \epsilon X/2}\rho e^{i \epsilon X/2}  + q X e^{-i \epsilon X/2}\rho e^{i \epsilon X/2}X \nonumber \\
&=& \Lambda_\epsilon \circ \Lambda_q [\rho], \label{error}
\end{eqnarray}
where $q$ is the probability of a stochastic bit-flip and $\epsilon$ is the angle of a small rotation error that is constant in time. We can relate these parameters to a physical dephasing rate $\gamma$ and systematic rotation at rate $\omega$ (e.g., from cross-talk or an external field) through the master equation
\begin{equation}
\frac{d \rho}{d t} = - i \frac{\omega}{2} [X,\rho] + \gamma(X\rho X - \rho),
\end{equation}
by setting $\epsilon = \omega \tau$ and $q = (1-e^{-2 \gamma \tau})/2$ for a gate time (QEC cycle time) $\tau$~\cite{combes_-situ_2014}.

Eq.~(\ref{error}) describes the composition of a coherent process, $\Lambda_\epsilon$, and an incoherent process, $\Lambda_q$. The latter is an appropriate description for environmentally induced decoherence as well as for random coherent rotations, such as those due to fluctuating control noise. These are described by the average over many instances of $U = e^{-i \theta X/2}$, where the angle $\theta$ fluctuates from one QEC cycle to the next. (Hence $q$ is the infidelity of the operator $U$, which can be related~\cite{nielsen_simple_2002} to the rms rotation angle as $q = \theta_{\mathrm{rms}}^2/4$.) Therefore $\Lambda_q$ and $\Lambda_\epsilon$ are suitable for describing the high and low-frequency components of a stochastic $X$ rotation error. Although this error model is somewhat restrictive in that the operators $\Lambda_q$ and $\Lambda_\epsilon$ commute, it captures the relevant impact of coherent errors on qubit error metrics~\cite{ball_effect_2016,kueng_comparing_2016}. 

We note that in general, it is possible to have a different rotation angle $\epsilon_j$ and a different bit-flip rate $q_j$ for each qubit. We are interested in capturing the properties of errors that have broad spatial extent such as external fields. It is therefore only important that these parameters have similar (non-zero) magnitude. Choosing them all identical as in Eq.~(\ref{error-full}) simplifies the calculations without sacrificing any significant generality. 

\subsection{Errors due to qubit-qubit interactions} \label{qubit-qubit-section}
Before presenting the analysis of single qubit errors, we wish to emphasize the importance of qubit-qubit interactions as a source of coherent error. Interactions contribute to coherent errors affecting data qubits and to syndrome errors when interactions occur between data and syndrome qubits. Generally, interactions entangle all the qubits and so are complicated to analyze. We do not consider them further beyond the example given in this section.

Here we comment on a special case that is simple but relevant, motivated by the hardware proposal in~\cite{chow_implementing_2014}. In this architecture, resonator-mediated qubit-qubit interactions are a dominant error source, resulting in an effective $ZZ$ interaction between the qubits that cannot be turned off~\cite{sheldon_procedure_2016}. We consider such an interaction between data qubits only (although in the architecture of~\cite{chow_implementing_2014} there will also be interactions between data and syndrome qubits), and show that it is fully correctable. 

For an interaction time $t$, the error is of the form
\begin{equation}
\mathcal{N}_{\mathrm{int}} = e^{-iH t} = \prod_{\langle i,j \rangle} e^{-iH_{ij}t}, \label{error-two-qubit}
\end{equation}
where the angular brackets indicate that the product is restricted to pairs of qubits connected by a resonator, and 
\begin{equation}
H_{ij} = \eta_{ij}Z_i\otimes Z_j. \label{hamiltonian-two-qubit}
\end{equation}
Here $\eta_{ij}$ is the interaction strength between qubits $i$ and $j$. The error operator, Eq.~(\ref{error-two-qubit}), factors into terms belonging to individual qubit pairs because the corresponding Hamiltonian terms, Eq.~(\ref{hamiltonian-two-qubit}), commute.

To make contact with our repetition code analysis in terms of $X$ errors, we now consider the error
\begin{equation}
\mathcal{N}_\mathrm{int} = \prod_{\langle i,j \rangle} e^{-i\epsilon_{ij}X_i\otimes X_j},
\end{equation}
where $\epsilon_{ij}$ is a generalized rotation angle. Expanding the exponential, we can write this error operator as a sum over all even-weight tensor products of $X$'s,
\begin{equation}
\mathcal{N}_\mathrm{int} = C^{(0)}+\sum_{i\neq j} C^{(2)}_{ij}X_i X_j + \sum_{i\neq j \neq k \neq l} C^{(4)}_{ijkl} X_i X_j X_k X_l + \ldots, \label{N-int}
\end{equation}
with some constants $C^{(2n)}_{ij...}$. (We have omitted the $\otimes$'s between $X$'s to save space.)

Each term in Eq.~(\ref{N-int}) corresponds to a unique syndrome. Since the code distance is odd, there is no redundancy in the syndromes that result from errors of weight less than or equal to $(d-1)/2$ and those with weight greater than $(d-1)/2$. (Compare Eq.~(\ref{Lambda-epsilon-otimes-N}) in the Appendix). We thus find the surprising result that coherent errors of the form Eq.~(\ref{N-int}) are uniquely projected to Pauli errors by stabilizer measurement and therefore are perfectly correctable. The caveat is that the decoding protocol must be modified from the standard one. For example, the syndrome normally corresponding to $X_1$ now indicates an error $X_2X_3\ldots X_N$.

This result illustrates the possibility of tailoring the decoding protocol to correct for a dominant error of a particular type. However, if there are multiple comparable error sources one runs into difficulties. In the present example, if there are single-qubit errors, Eqs.~(\ref{error-full})-(\ref{error}), the decoding protocol tailored to $ZZ$-interactions gives a logical error, so these errors are amplified by a factor of order $N$. Therefore this approach is only useful if the rate of $ZZ$-interaction errors is much greater than $Np$, where $p$ is the single-qubit error probability.

\section{Analysis}
Upon logical encoding with the repetition code, we obtain an effective error model for the logical qubit that has the same form as Eq.~(\ref{error}) but with renormalized parameters $q$ and $\epsilon$, as shown in the Appendix. This mirrors the general transformation found in~\cite{fern_generalized_2006}, and stems from the fact that unital and trace preserving physical errors lead to unital, trace preserving logical errors. Such channels can be expressed as Pauli times a unitary. Non-unital deviations from Pauli channels were found in~\cite{wallman_bounding_2015} to not change the error rate significantly so we do not consider them here. 

For low physical error rates, $q, |\epsilon| \ll 1$, we find the following recursion relations giving the values of $q$ and $\epsilon$ at the $(n+1)$-st level of concatenation in terms of the ones at level $n$.
\begin{eqnarray}
\epsilon_{n+1} &=& 2 \binom{d-1}{(d-1)/2} \left( \frac{\epsilon_n}{2}\right)^d, \label{epsilon-recursion} \\
q_{n+1} &=& \binom{d}{(d-1)/2} \left(\frac{\epsilon_n^2}{4} + q_n\right)^{(d+1)/2}. \label{q-recursion}
\end{eqnarray}
Here $d=N$ is the code distance. The computation is elementary but lengthy. Details are given in the Appendix.

These recursion relations saturate the bounds given in ~\cite{fern_generalized_2006}. In terms of our parameters, these bounds are: $\epsilon_{n+1}$ is at most $\mathcal{O}(\epsilon_n^d)$ and $q_{n+1}-q_{n+1}|_{\epsilon_n = 0}$ is at most $\mathcal{O}(\epsilon_n^2)$. The second of these bounds is not tight when $q_n = 0$, and so cannot be used to determine the impact of the coherent error at level $n+1$ on the logical error rate.

\subsection{Coherent errors and code concatenation}
\begin{figure}
\includegraphics[width=84mm]{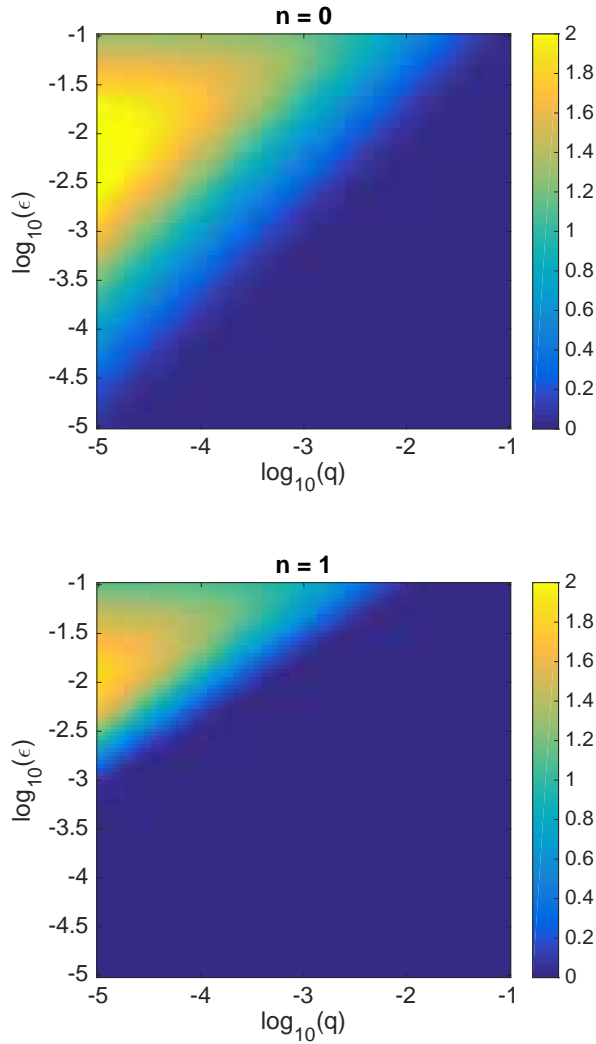}
\caption{Coherence of the error, $\log_{10}\left(\frac{2}{3}\frac{D}{r}\right)$, for the physical qubit (concatenation level $n=0$) and logical qubit ($n=1$) of the distance 3 repetition code. Larger values indicate greater coherence. Both plots refer to the same range of initial values of $\epsilon$ and $q$ in Eq.~(\ref{error}) for the physical error. For higher concatenation levels, $n\geq 2$ (not shown), the ratio $D/r$ does not differ significantly from $3/2$ anywhere in the given parameter region, indicating that the error is completely incoherent in this case. \label{D-over-r}}
 \end{figure}

We are interested in the extent to which the effective error at concatenation level $n$ is coherent, and at what value of $n$ this error becomes effectively stochastic. As discussed in~\cite{kueng_comparing_2016}, a convenient metric for doing so is the ratio $D/r$ of the diamond distance $D$~\cite{KitaevBook} to the average infidelity $r$~\cite{horodecki_general_1999,nielsen_simple_2002} of the channel. For the error channel, Eq.~(\ref{error}), these quantities are~\cite{kueng_comparing_2016}
\begin{eqnarray}
r_n &=& \frac{2}{3}\left[q_n\cos(\epsilon_n) + \sin^2(\epsilon_n/2)\right], \label{rn} \\
D_n &=& \sqrt{\frac{3}{2}r_n - q_n(1-q_n)}. \label{Dn}
\end{eqnarray}
For the purely incoherent case, $\epsilon_n=0$, we have $D_n = q_n = \frac{3}{2}r_n$. Therefore the ratio $D_n/r_n$ should tend to $3/2$ from above as the coherent contribution becomes negligible. 

In the physically relevant regime of small initial values $q, |\epsilon| \ll 1$, $q_n$ and $\epsilon_n$ are also small. Eqs.~(\ref{rn})-(\ref{Dn}) then give
\begin{equation}
\frac{D_n}{r_n} \approx \frac{3}{2}\frac{\sqrt{q_n^2 + \epsilon_n^2/4}}{q_n + \epsilon_n^2/4}.\label{Dn-rn}
\end{equation}
Therefore the error channel is incoherent if $|\epsilon_n| \ll q_n$. Using the recursion relations, Eqs.~(\ref{epsilon-recursion})-(\ref{q-recursion}), it is straightforward to show that this occurs when both $n \ge 2$ and $d \ge 3$ for any initial $q,|\epsilon| \ll 1$. Hence only two levels of concatenation are necessary to obtain a stochastic effective logical error channel, for any size repetition code.

\begin{figure}
\includegraphics[width=84mm]{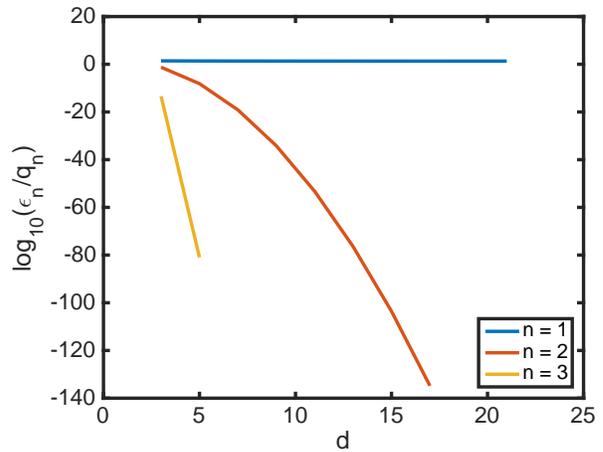}
\caption{The coherence metric, $\epsilon_n/q_n$, calculated from Eqs.~(\ref{epsilon-recursion})-(\ref{q-recursion}), vs code distance $d$. The values shown are for the initial condition $(q,\epsilon) = (0,0.1)$ in Eq.~(\ref{error}) and concatenation levels $n=1,2,3$. When $n\geq 2$ and $d \geq 3$, the effective logical channel is incoherent, $|\epsilon_n|/q_n \ll 1$. \label{q-over-epsilon}}
 \end{figure}

Figs.~\ref{D-over-r} and~\ref{q-over-epsilon} illustrate this result. Fig.~\ref{D-over-r} plots the ratio of Eq.~(\ref{Dn}) and Eq.~(\ref{rn}) (without approximation) for $d=3$, $n=0,1$ and a range of initial values of $\epsilon$ and $q$. Essentially no deviation from $D/r=3/2$ is observed when $n\geq 2$. Fig.~\ref{q-over-epsilon} plots $\epsilon_n/q_n$ vs $d$ for a single initial value $\epsilon = 0.1$ and $q=0$, and for several values of $n$. From Eqs.~(\ref{epsilon-recursion})-(\ref{q-recursion}) it is clear that $|\epsilon_n|/q_n$ for any initial condition $(q,\epsilon)=(q_0,\epsilon_0)$ is upper-bounded by its value for $(q,\epsilon) = (0,\epsilon_0)$. The plot shows that, for a given set of initial parameters, the encoded channel becomes more incoherent with increasing $d$.

\subsection{Logical time to failure}

We now examine the logical time to failure of the encoded qubit as a function of concatenation level, $n$. As a failure metric we use the worst-case infidelity~\cite{wallman_bounding_2015} after $m$ QEC cycles, defined as 
\begin{equation}
w_n(m) \equiv \max_{|\psi_n(0)\rangle} \left[1-|\langle\psi_{n}(0)|\psi_{n}(m)\rangle|^2 \right]. \label{infidelity}
\end{equation}
This is the maximum probability of logical failure over all initial states. Here $|\psi_{n}(m)\rangle$ is the state of the level-$n$ concatenated logical qubit after $m$ QEC cycles,
\begin{equation}
|\psi_{n}(m)\rangle=\underbrace{\mathcal{N}_n\circ \cdots \circ \mathcal{N}_n}_{m \text{ times}}\left(|\psi_{n}(0)\rangle\right). 
\end{equation}
The noise operator $\mathcal{N}_n$ is the single qubit operator, Eq.~(\ref{error}), with $\epsilon$, $q$ replaced by $\epsilon_n$, $q_n$ from Eqs.~(\ref{epsilon-recursion})-(\ref{q-recursion}).
Eq.~(\ref{infidelity}) is satisfied for some initial state $|\psi_n(0)\rangle=|\psi_n^*(0)\rangle$. For the noise operator $\mathcal{N}_n$, $|\psi_{n}^*(0)\rangle$ is any state of the form $\cos(\theta)|0\rangle - i\sin(\theta)| 1\rangle$. This is a state whose Bloch vector is in the $y$--$z$ plane and is therefore maximally affected by the $X$-rotation in Eq.~(\ref{error}). 

Since the worst-case infidelity of the error channel, Eq.~(\ref{error}), is $3/2$ the average infidelity, we find from Eq.~(\ref{rn}) that
\begin{equation}
w_m^{(n)} = mq_n\cos(m\epsilon_n) + \sin^2\left(\frac{m\epsilon_n}{2}\right). \label{infidelity-level-n}
\end{equation}
In the limit $q_n, |\epsilon_n| \ll 1/m$, this becomes
\begin{equation}
w_m^{(n)} \approx mq_n + \left(\frac{m\epsilon_n}{2}\right)^2. \label{infidelity-level-n-approx}
\end{equation}
This is a sum of two error rates, the first from the incoherent channel $\Lambda_{q_n}^m = \Lambda_{mq_n}$ and the second from the coherent rotation, $\Lambda_{\epsilon_n}^m = \Lambda_{m\epsilon_n}$. (The notation $\Lambda^m$ denotes the composition of $\Lambda$ with itself $m$ times.)

Eq.~(\ref{infidelity-level-n}) predicts logical failure at $m_{\mathrm{fail}} = \mathcal{O}(1/{\mathrm{max}(\epsilon_n,q_n)})$ QEC cycles. This is the same scaling one obtains using the diamond distance as an error metric. Indeed, Eq.~(\ref{Dn}) gives the diamond distance after $m$ QEC cycles as
\begin{equation}
D_n(m) = m D_n(1) = m\sqrt{q_n^2 +\frac{\epsilon_n^2}{4}},
\end{equation} 
from which the result follows by setting $D_n(m_{\mathrm{fail}}) \sim 1$. 

Eq.~(\ref{infidelity-level-n}) also predicts a crossover from stochastic behavior, $w_m \approx mq_n$, to coherent behavior, $w_m \approx \sin^2(m\epsilon_n/2)$, above a critical number $m_{\mathrm{crit}}$ of QEC cycles,
\begin{equation}
m_{\mathrm{crit}} \sim \frac{2q_n}{\epsilon_n^2}.\label{m-crit}
\end{equation} 
We showed above that $\epsilon_n \ll q_n$ for $n \geq 2$. Therefore the logical failure rate has a coherent contribution only for $n=1$. In this case, inserting the values of $\epsilon_1$, $q_1$ from Eqs.~(\ref{epsilon-recursion})-(\ref{q-recursion}) into Eq.~(\ref{m-crit}) we find that the crossover from stochastic to coherent behavior of the logical error is lower-bounded by $m_{\mathrm{crit}} \sim 1/\epsilon^{d-1}$, which holds when $q=0$. In contrast, the number of QEC cycles to logical failure at $n=1$ is lower-bounded by  $m_{\mathrm{fail}} \sim 1/\epsilon^d$, which is $\mathcal{O}(1/\epsilon)$ cycles greater than the crossover point.

For comparison, the stochastic error model defined by the Pauli-twirl of Eqs.~(\ref{error-full})-(\ref{error}) gives $\epsilon_1=0$ in Eq.~(\ref{epsilon-recursion}). The worst-case logical failure probability is then equal to the first term in Eq.~(\ref{infidelity-level-n-approx}), which is approximately the stochastic part of the full error probability. This model predicts a logical time to failure proportional to $m_{\mathrm{fail}}^{\mathrm{st}} = \mathcal{O}(1/\epsilon^{d+1})$. This is $\mathcal{O}(1/\epsilon)$ longer (more QEC cycles) than when the coherent part of the error was included. 

We conclude that a stochastic error model would correctly predict the logical failure rate up to $m_{\mathrm{crit}}$ QEC cycles, beyond which it is necessary to take account of the coherence of the error. If the correlation time of the coherent error is less than $\tau_{\mathrm{coh}} = m_{\mathrm{crit}}\tau_{QEC}$, where $\tau_{QEC}$ is the duration of a QEC cycle, then the logical error is effectively stochastic and can be obtained by replacing the physical error, Eq.~(\ref{error}), by its Pauli twirl.

\begin{figure}
\includegraphics[width=84mm]{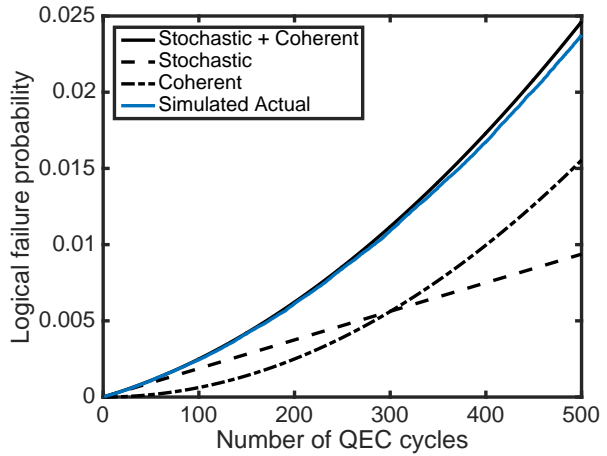}
\caption{Logical failure probability of the 3-qubit repetition code with coherent errors given by Eqs.~(\ref{error-full})-(\ref{error}) for $n=1$ concatenation level. The rotation angle is $\epsilon = 0.1$ radians and the depolarizing probability is $q=0$. The vertical axis is the worst-case logical error probability, defined in Eq.~(\ref{infidelity}) and the accompanying text. The blue line is the average of 10,000 Monte Carlo sample runs with data qubits initially in state $|000\rangle$. The error operator is applied once per QEC round and syndrome extraction is perfect. The black lines are theory curves where ``Stochastic" is the first term in Eq.~(\ref{infidelity-level-n-approx}), ``Coherent" is the second term in this equation, and ``Stochastic + Coherent" is Eq.~(\ref{infidelity-level-n}), all with $n=1$. The simulation and theory show good agreement. The stochastic approximation begins to fail around $1/\epsilon^2 = 100$ QEC cycles, consistent with the discussion in the text. \label{failure-prob-fig}}
\end{figure}

Fig.~\ref{failure-prob-fig} plots the worst-case failure probability vs number of QEC cycles for physical error parameters $\epsilon = 0.1$, $q=0$ and for $d=3$. The blue curve is the result of a Monte Carlo simulation of three data qubits initialized to $|\psi_{1}^*(0)\rangle = |0\rangle$, each subject to the error operator, Eq.~(\ref{error}), once per QEC cycle. This is compared to the theoretical curve, Eq.~(\ref{infidelity-level-n}), as well as the coherent and incoherent parts of Eq.~(\ref{infidelity-level-n-approx}). The simulation results show good agreement with Eq.~(\ref{infidelity-level-n}). The crossover from stochastic to coherent behavior occurs at $m_{\mathrm{crit}} \sim 1/\epsilon^2 = 100$, consistent with the discussion above.

\section{Conclusion}

Implementing quantum error correction for coherent errors will be an important challenge for realizing scalable fault-tolerant quantum computing. It is now understood that despite the projective nature of QEC stabilizer measurements, coherent physical errors give rise to a logical error that is also coherent to some extent~\cite{fern_generalized_2006,gutierrez_errors_2016,darmawan_tensor-network_2016}. Several strategies for mitigating coherent errors have been proposed, including the realization of Pauli twirling through random updates to the Pauli frame~\cite{kern_quantum_2005,knill_quantum_2005,wallman_noise_2016}, and optimization of the decoding algorithm~\cite{chamberland_hard_2016}. Our goal in this paper was to investigate another possibility, namely whether there exist parameter regimes in standard stabilizer QEC for which the logical error is Pauli even when the physical error is coherent. This would enable the use of existing techniques for correcting Pauli errors, and justify numerical simulations of QEC based on a Pauli error model.

To this end, we analyzed repetition code quantum error correction for an error model containing coherent and incoherent errors. Focusing on the repetition code allowed us to obtain quantitative analytic results on the scaling of coherent and incoherent logical error rates with code distance and concatenation. 

We found that coherent physical errors result in logical errors that are partially coherent and therefore non-Pauli, in agreement with recent numerical studies~\cite{gutierrez_errors_2016,darmawan_tensor-network_2016}, but that the degree of coherence depends on the code distance and concatenation level. At one concatenation level, the coherent part of the logical error is negligible at fewer than $\epsilon^{-(d-1)}$ error correction cycles, where $\epsilon \ll 1$ is the rotation angle error per cycle for a single physical qubit and $d$ is the code distance. Logical failure occurs at $\mathcal{O}(1/\epsilon^d)$ QEC cycles, which is $\mathcal{O}(1/\epsilon)$ faster than predicted by the Pauli-twirl approximation of the error.  At two or more concatenation levels, the coherent part of the logical error is negligible at any number of error correction cycles for small initial error rates and any distance $d$.

The result on concatenation is worth emphasizing. It is remarkable that two concatenation levels suffice to make the logical error effectively stochastic over a broad range of initial error rates (See Fig.~\ref{D-over-r}). A similar result holds for the $d=3$ Steane code, as can be shown by inserting our error model, Eqs.~(\ref{error-full})-(\ref{error}), in the general equations of Ref.~\cite{fern_generalized_2006}. If true for general codes, this opens the possibility of using code concatenation to eliminate coherent errors.

Our findings lend support to using stochastic Pauli error models in the presence of low-frequency coherent errors under conditions of large enough code distance or concatenation. However, several questions remain to be addressed to justify the use of Pauli models for general coherent errors. Our error model was limited to a tensor product of single-qubit errors on data qubits only. Future work must include syndrome qubit and measurement errors, coherent leakage errors~\cite{wallman_robust_2016,kueng_comparing_2016}, and gate dependence. In addition, recent hardware proposals (e.g.,~\cite{chow_implementing_2014}) include couplings both between data and syndrome qubits as well as between pairs of datas and syndromes. Such couplings cause coherent syndrome errors that need to be included in the analysis.

Finally, while the calculations presented here are suggestive, the real test lies in their applicability to the particular QEC codes that will be implemented in real devices. To this end, we hope these results may inform further studies of, e.g., the surface code, which at present can only be done numerically.

\section{Acknowledgements}
We are grateful to Marcus P. da Silva for valuable discussions. This document does not contain technology or technical data controlled under either U.S. International Traffic in Arms Regulations or the U.S. Export Administration Regulations.

\appendix*
\section{Derivation of the logical error map}
\label{appendix}

We derive the recursion relations, Eqs.~(\ref{epsilon-recursion})-(\ref{q-recursion}), for the repetition code under the error model, Eqs.~(\ref{error-full})-(\ref{error}). We follow the notation of Rahn, {\it{et al.}}~\cite{rahn_exact_2002}. The effective logical error map is the composition of encoding, noise, and decoding,
\begin{equation}
\mathcal{G}=\mathcal{D}\circ\mathcal{N}\circ\mathcal{E}.
\end{equation}
The encoding map for the repetition code is 
\begin{equation}
\mathcal{E}: |0\rangle \mapsto |\bar{0}\rangle = |00\cdots 0\rangle, |1\rangle \mapsto |\bar{1}\rangle = |11\cdots 1\rangle.
\end{equation}

The decoding map includes syndrome measurement and recovery, followed by the inverse of encoding. It can be expressed as a sum over all syndromes,
\begin{equation}
\mathcal{D}=\sum_\sigma \mathcal{E}^\dag\circ\mathcal{R}_\sigma\circ\mathcal{P}_\sigma,
\end{equation}
where
\begin{equation}
\mathcal{P}_\sigma = \prod_i \frac{1+\sigma_i S_i}{2}
\end{equation}
is the projection operator corresponding to the syndrome $\sigma=(\sigma_1,\cdots,\sigma_{N-1})$, $S_i=Z_i Z_{i+1}$ is the $i$-th stabilizer, and $\mathcal{R}_\sigma$ is the recovery operation that maps the logical state back to the codespace.

To simplify the notation, we work directly in terms of the physical qubit density matrix, $\bar{\rho} = \mathcal{E}(\rho)$ and drop the $\mathcal{E}$ operators. The logical error map is then given by $\mathcal{G} = \mathcal{E}^\dag\circ\bar{\mathcal{G}}\circ\mathcal{E}$, where
\begin{equation}
\bar{\mathcal{G}} = \sum_\sigma \mathcal{R}_\sigma\circ\mathcal{P}_\sigma\circ\mathcal{N}_N(\bar{\rho}).
\end{equation}

We now factor the stochastic and coherent parts of $\mathcal{N}_N$:
\begin{equation}
\mathcal{N}_N = \Lambda_q^{\otimes N}\circ\Lambda_\epsilon^{\otimes N}.\label{N}
\end{equation}
The incoherent part can be written
\begin{eqnarray}
\Lambda_q^{\otimes N} &=& \sum_{k=0}^N q^k(1-q)^{N-k} \mathcal{O}_k \nonumber \\
&=& \sum_{k=0}^{(N-1)/2} \mathcal{O}_k \left[q^k(1-q)^{N-k} +q^{N-k}(1-q)^k \bar{X}\right].\nonumber \\ \label{Lambda-q-N}
\end{eqnarray}
Here $\mathcal{O}_k$ is the sum over all weight-$k$ products of $X_i$'s:
\begin{equation}
\mathcal{O}_k = X_1 X_2 \cdots X_k + X_1 X_2 \cdots X_{k-1}X_{k+1} + \cdots.
\end{equation}
The operator $\bar{X} = \mathcal{O}_N$ is the lowest-weight undetectable error, and is also the logical bit-flip operator~\cite{raussendorf_key_2012}. The second line of Eq.~(\ref{Lambda-q-N}) comes from the fact that $\mathcal{O}_j = \bar{X}\mathcal{O}_{N-j}$.

We can write the coherent part of Eq.~(\ref{N}) as
\begin{equation}
\Lambda_\epsilon^{\otimes N} = \bigotimes_{j=1}^N (c - i s X_j),
\end{equation}
where
\begin{eqnarray}
c &\equiv& \cos\left(\frac{\epsilon}{2}\right), \\
s &\equiv& \sin\left(\frac{\epsilon}{2}\right).
\end{eqnarray}
This can be expanded as
\begin{equation}
\Lambda_\epsilon^{\otimes N} = \sum_{j=0}^{(N-1)/2}[c^{N-j}(-i s)^j + c^j (-i s)^{N-j}\bar{X}]\mathcal{O}_j. \label{Lambda-epsilon-otimes-N}
\end{equation}
We note that the error $\Lambda_\epsilon^{\otimes N}$ includes coherent linear combinations of correctable and uncorrectable errors. The term in brackets defines a unitary operator,
\begin{equation}
\bar{U}_j \equiv \frac{c^{N-j}(-i s)^j + c^j (-i s)^{N-j}\bar{X}}{\sqrt{c^{2(N-j)} s^{2j} + c^{2j}s^{2(N-j)}}}.
\end{equation}
This can be written
\begin{equation}
\bar{U}_j  = \exp\left(-i\frac{\epsilon_j}{2}\bar{X}\right),
\end{equation}
where the rotation angle $\epsilon_j$ is given by
\begin{equation}
\tan\left(\frac{\epsilon_j}{2}\right) = (-1)^{\frac{N-1}{2}+j}\tan^{N-2j}\left(\frac{\epsilon}{2}\right). \label{epsilon-j}
\end{equation}

Collecting terms, we can write the full error map, Eq.~(\ref{N}), as
\begin{eqnarray}
\mathcal{N}_N = &\sum_{j,k=0}^{(N-1)/2}&P_j[q^k(1-q)^{N-k}+ q^{N-k}(1-q)^k \bar{X}]\nonumber\\
&&\circ\bar{U}_j\circ\mathcal{O}_j\circ\mathcal{O}_k, \label{NN}
\end{eqnarray}
where $P_j$ is the probability of $\bar{U}_j$,
\begin{equation}
P_j = c^{2(N-j)} s^{2j} + c^{2j}s^{2(N-j)}.\label{Pj}
\end{equation}

Since each term in the product $\mathcal{O}_j \circ \mathcal{O}_k$ is a product of $X_i$'s, it follows that
\begin{equation}
\mathcal{O}_j \circ \mathcal{O}_k = \sum_{n=0}^{\min(j,k)} m_{jk}^n \mathcal{O}_{k+j-2n} \label{OjOk}
\end{equation}
for some coefficients $m_{jk}^n$. 

It is straightforward to calculate these coefficients. Consider a single term in $\mathcal{O}_k$ and select $n$ of the $X_i$'s in this term to coincide with terms in $\mathcal{O}_j$. The appropriate terms in $\mathcal{O}_j$ are chosen by selecting from $(N-n) - (k-n) = N-k$ possible $X_i$'s for the remaining $j-n$ $X_i$'s in $\mathcal{O}_j$ that contribute to the RHS of Eq.~(\ref{OjOk}). This gives $N-k \choose j-n$ terms. Next there are $k \choose n$ ways to pick $n$ $X_i$'s in the given term of $\mathcal{O}_k$. Finally there are $N \choose k$ terms in $\mathcal{O}_k$. Since each term on the LHS of Eq.~(\ref{OjOk}) contributes to the RHS of Eq.~(\ref{OjOk}) and there are $N \choose j+k-2n$ terms on the RHS, we obtain
\begin{equation}
m_{jk}^n = {N\choose k}{k\choose n}{N - k \choose j - n} / {N \choose k+j-2n}.\label{mjkn}
\end{equation}

Substituting Eq.~(\ref{OjOk}) into Eq.~(\ref{NN}) gives
\begin{eqnarray}
\mathcal{N}_N &=& \sum_{j,k=0}^{(N-1)/2}\sum_{n=0}^{\min(j,k)}m_{jk}^nP_j\bar{U}_j\circ\mathcal{O}_{k+j-2n}\nonumber\\
&&\circ[q^k(1-q)^{N-k}+ q^{N-k}(1-q)^k \bar{X}].
\end{eqnarray}
Now the only operator in this equation that creates a non-trivial syndrome is $\mathcal{O}_{k+j-2n}$. If $k+j-2n \leq (N-1)/2$ then the recovery operation transforms each of the $N \choose k+j-2n$ terms in this operator to the identity, while if $k+j-2n > (N-1)/2$ then the recovery operation transforms each term in this operator to $\bar{X}$. Substituting Eq.~(\ref{mjkn}) for $m_{jk}^n$ and collecting terms, we find
\begin{equation}
\bar{\mathcal{G}}(\bar{\rho}) = \sum_{j=0}^{(N-1)/2} {N \choose j} P_j(A_j\bar{U}_j\bar{\rho}\bar{U}_j^\dag + Q_j \bar{X}\bar{U}_j\bar{\rho}\bar{U}_j^\dag\bar{X}),
\end{equation}
where
\begin{widetext}
\begin{eqnarray}
A_j &=& \sum_{k=0}^{(N-1)/2}\sum_n {j \choose n}{N-j \choose k-n}\left[\Theta\left(j+k-2n \leq \frac{N-1}{2}\right)q^k(1-q)^{N-k} +  \Theta\left(j+k-2n > \frac{N-1}{2}\right)q^{N-k}(1-q)^k\right],\nonumber \\ \\
Q_j &=& \sum_{k=0}^{(N-1)/2}\sum_n {j \choose n}{N-j \choose k-n}\left[\Theta\left(j+k-2n \leq \frac{N-1}{2}\right)q^{N-k}(1-q)^k +  \Theta\left(j+k-2n > \frac{N-1}{2}\right)q^k(1-q)^{N-k}\right]. \nonumber \\ \label{Qj}
\end{eqnarray}
\end{widetext}
The sum over $n$ runs over all values for which the combinatorial coefficients are defined. Here $\Theta(x)$ is a Heaviside step function which takes the value 1 when the statement $x$ is true and 0 when it is false.

Using Vandermonde's identity,
\begin{equation}
\sum_n {j \choose n}{N-j \choose k-n} = {N \choose k}, 
\end{equation}
we find
\begin{equation}
A_j + Q_j = 1.
\end{equation}
Returning now to the full error map $\mathcal{G} = \mathcal{E}^\dag\circ\bar{\mathcal{G}}\circ\mathcal{E}$ we obtain for the effective 1-qubit logical error channel
\begin{equation}
\mathcal{G} = \sum_{j=0}^{(N-1)/2} {N \choose j} P_j \Lambda_{Q_j}\circ\Lambda_{\epsilon_j}, \label{G}
\end{equation}
where
\begin{eqnarray}
\Lambda_{Q_j}(\rho) &=& (1-Q_j)\rho + Q_j X\rho X,\label{Lambda-Qj}\\
\Lambda_{\epsilon_j}(\rho) &=& U_j \rho U_j^\dag.\label{Lambda-epsilonj}
\end{eqnarray}

Eqs.~(\ref{G})-(\ref{Lambda-epsilonj}) show that the effective logical error is an incoherent sum of terms, each of which is the composition of a stochastic bit flip and coherent rotation.

To enable further simplification it is helpful to use the Liouville, or Pauli Transfer Matrix (PTM) representation of quantum channels~\cite{havel_robust_2003}. For an $N$-qubit channel this is defined as
\begin{equation}
R_{ij}(\Lambda) = \frac{1}{2^N}\mathrm{Tr}\left[V_i\Lambda(V_j)\right],
\end{equation}
where $V_i$, $V_j \in \{1,X,Y,Z\}^{\otimes N}$ are basis vectors in the vector space spanned by all tensor products of $N$ Pauli matrices (including the identity).

The PTMs are linear functions of their arguments, and map composition is represented by matrix multiplication. The following identities hold for arbitrary quantum channels $\Lambda_1$, $\Lambda_2$ and constants $a$, $b$.
\begin{eqnarray}
R_{ij}(\Lambda_1\circ\Lambda_2) &=& \sum_k R_{ik}(\Lambda_1)R_{kj}(\Lambda_2), \label{R-mult} \\
R_{ij}(a\Lambda_1 + b\Lambda_2) &=& aR_{ij}(\Lambda_1) + bR_{ij}(\Lambda_2). \label{R-add}
\end{eqnarray}

The logical error channel, Eq.~(\ref{G}), is a single-qubit channel. It contains only identity and $X$ operators, leaving the space spanned by $\{1,X\}$ invariant. By analogy to the notation in Ref.~\cite{gutierrez_errors_2016} for process matrices, we can therefore write $R$ in terms of 2 X 2 blocks,
\begin{equation}
R = \left(\begin{array}{cc} 1 & 0 \\ 0 & \tilde{R} \end{array}\right).
\end{equation}
The matrices $\tilde{R}$ satisfy the same composition properties, Eqs.~(\ref{R-mult})-(\ref{R-add}) as the full matrix $R$.

We now write Eq.~(\ref{G}) as
\begin{eqnarray}
\tilde{R}(\mathcal{G}) &=& \sum_{j=0}^{(N-1)/2} {N \choose j} P_j \tilde{R}(\Lambda_{Q_j})\tilde{R}(\Lambda_{\epsilon_j})\\
&=&  \sum_{j=0}^{(N-1)/2} {N \choose j} P_j \left(1-2Q_j\right)\left(\begin{array}{cc} \cos\epsilon_j & -\sin\epsilon_j \\ \sin\epsilon_j & \cos\epsilon_j \end{array}\right). \nonumber \\ \label{R-tilde}
\end{eqnarray}

We are interested in the leading order terms in this equation in $\epsilon$, $q$. From Eq.~(\ref{Pj}) we find to lowest order,
\begin{equation}
P_j \approx \left\{\begin{array}{ll} 1 - \sum_{k=1}^{(N-1)/2}{N \choose k}\left(\frac{\epsilon}{2}\right)^{2k}, & j=0, \\ \\
\left(\frac{\epsilon}{2}\right)^{2j}, & j \neq 0, \end{array}\right.
\end{equation}
and from Eq.~(\ref{Qj}) we find
\begin{equation}
Q_j \approx {N-j \choose \frac{N+1}{2} - j} q^{\frac{N+1}{2}-j}.
\end{equation}
The last equation is found by observing that the lowest order term in $q$ comes from the 2nd term in Eq.~(\ref{Qj}) when $k$ takes its minimal value. This happens when $n=0$ and $k=(N+1)/2-j$. Finally, from Eq.~(\ref{epsilon-j}) we find
\begin{equation}
\epsilon_j \approx \left(\frac{i}{2}\right)^{N-2j-1}\epsilon^{N-2j}.
\end{equation}

We now calculate the diagonal and off-diagonal terms in Eq.~(\ref{R-tilde}). The diagonal terms are
\begin{widetext}
\begin{eqnarray}
\sum_{j=0}^{(N-1)/2} {N \choose j} P_j \left(1-2Q_j\right) \cos\epsilon_j &=& \left[1 - \sum_{k=1}^{(N-1)/2}{N \choose k}\left(\frac{\epsilon}{2}\right)^{2k}\right]
\left[1-2 {N\choose \frac{N+1}{2}} q^{\frac{N+1}{2}-j}\right] \left[1-2\left(\frac{\epsilon}{2}\right)^{2N} \right] \nonumber \\
&&+ \sum_{j=1}^{(N-1)/2} {N \choose j} \left(\frac{\epsilon}{2}\right)^{2j} \left[1-2 {N-j \choose \frac{N+1}{2} - j} q^{\frac{N+1}{2}-j}\right] \left[1-2\left(\frac{\epsilon}{2}\right)^{2N-4j} \right].
\end{eqnarray}
\end{widetext}
Expanding and keeping the lowest order terms in $q$ and $\epsilon$, we find that the above equation reduces to $1-2\bar{q}$, where
\begin{equation}
\bar{q} = {N \choose \frac{N+1}{2}} \left(\frac{\epsilon^2}{4} + q\right)^\frac{N+1}{2}. 
\end{equation}
This proves Eq.~(\ref{q-recursion}).

Next we calculate the off-diagonals in Eq.~(\ref{R-tilde}).
\begin{widetext}
\begin{eqnarray}
\sum_{j=0}^{(N-1)/2} {N \choose j} P_j \left(1-2Q_j\right) \sin\epsilon_j &=& \left[1 - \sum_{k=1}^{(N-1)/2}{N \choose k}\left(\frac{\epsilon}{2}\right)^{2k}\right]
\left[1-2 {N\choose \frac{N+1}{2}} q^{\frac{N+1}{2}-j}\right] \left[2\left(-1\right)^\frac{N-1}{2}\left(\frac{\epsilon}{2}\right)^{N} \right] \nonumber \\
&&+ \sum_{j=1}^{(N-1)/2} {N \choose j} \left(\frac{\epsilon}{2}\right)^{2j} \left[1-2 {N-j \choose \frac{N+1}{2} - j} q^{\frac{N+1}{2}-j}\right] \left[2\left(-1\right)^{\frac{N-1}{2}-j}\left(\frac{\epsilon}{2}\right)^{N-2j} \right] \nonumber \\
&\approx& 2 \left(\frac{\epsilon}{2}\right)^{N} \sum_{j=0}^{(N-1)/2}\left(-1\right)^{\frac{N-1}{2}-j} {N \choose j}.
\end{eqnarray}
\end{widetext}
The sum in the last equation is straightforward to simplify by writing the sums over even and odd $j$ separately and using Pascal's rule. The result is
\begin{equation}
\bar{\epsilon} = 2 { N-1 \choose \frac{N-1}{2} } \left(\frac{\epsilon}{2}\right)^N.
\end{equation}
This proves Eq.~(\ref{epsilon-recursion}).

At this level of approximation (linear in $\bar{q}$ and $\bar{\epsilon}$), we can write the reduced PTM as
\begin{equation}
\tilde{R}(\mathcal{G}) = (1-2\bar{q})\left(\begin{array}{cc} \cos\bar{\epsilon} & -\sin\bar{\epsilon} \\ \sin\bar{\epsilon} & \cos\bar{\epsilon} \end{array}\right), \label{R-tilde-G}
\end{equation}
This shows that $\mathcal{G}$ has the same form as $\mathcal{N}$ [see Eq.~(\ref{error})] with renormalized parameters $q \rightarrow \bar{q}$, $\epsilon \rightarrow \bar{\epsilon}$.

\bibliography{mybib}

\end{document}